\documentclass[aps,twocolumn,showpacs,superscriptaddress,10pt]{revtex4-1}
\usepackage[utf8]{inputenc}
\usepackage{graphicx,amsmath,color}
\usepackage{amssymb}
\usepackage{amsthm}
\usepackage{times}
\usepackage{bbm}
\usepackage{bbold}
\usepackage[colorlinks,bookmarks=true,citecolor=blue,linkcolor=red,urlcolor=blue]{hyperref}

\newcommand{\je}[1]{\textcolor{black}{#1}}

\begin{document}

\title{Dissipative preparation of Chern insulators}

\author{Jan Carl Budich}

\affiliation{Institute for Theoretical Physics, University of Innsbruck, 6020 Innsbruck, Austria}
\affiliation{Institute for Quantum Optics and Quantum Information, Austrian Academy of Sciences, 6020 Innsbruck, Austria}

\author{Peter Zoller}
\affiliation{Institute for Theoretical Physics, University of Innsbruck, 6020 Innsbruck, Austria}
\affiliation{Institute for Quantum Optics and Quantum Information, Austrian Academy of Sciences, 6020 Innsbruck, Austria}

\author{Sebastian Diehl}
\affiliation{Institute for Theoretical Physics, University of Innsbruck, 6020 Innsbruck, Austria}
\affiliation{Institute for Quantum Optics and Quantum Information, Austrian Academy of Sciences, 6020 Innsbruck, Austria}
\affiliation{Institute of Theoretical Physics, TU Dresden, D-01062 Dresden, Germany}

\date{\today}
\begin{abstract}
Engineered dissipation can be employed to prepare interesting quantum many body states in a non-equilibrium fashion.
The basic idea is to obtain the state of interest as the unique steady state of a quantum master equation, irrespective of the initial state. Due to a fundamental competition of topology and locality, the dissipative preparation of gapped topological phases with a non-vanishing Chern number has so far remained elusive.
Here, we study the open quantum system dynamics of fermions on a two-dimensional lattice in the framework of a Lindblad master equation. In particular, we discover a mechanism to dissipatively prepare a topological steady state with non-zero Chern number by means of {\emph{short-range}} system bath interaction. Quite remarkably, this gives rise to a stable topological phase in a non-equilibrium phase diagram. We demonstrate how our theoretical construction can be implemented in a microscopic model that is experimentally feasible with cold atoms in optical lattices.
\end{abstract}
\maketitle

\section{Introduction}
The possibility of \emph{dissipatively}
preparing interesting quantum states of matter in open quantum systems \cite{poyatos96,DiehlDissPrep,Verstraete2008,Weimer2010,barreiroNature,Krauter2011,kapit14}
provides a new paradigm in quantum engineering complementing both practically and conceptually ongoing efforts on the implementation of many-body
Hamiltonians in closed quantum systems \cite{ZwergerReview,LewensteinReview,ColdFermionReview,jotzu14,aidelsburger14}.
The guiding idea of the dissipative approach is to engineer a controlled interaction of a quantum system with its environment in order to realize an exotic state of quantum matter in  a non-equilibrium fashion as the {\emph{unique}} steady state of a Markovian quantum master equation, irrespective of the initial state. The degree of control over the system bath interaction determines the precision of the dissipative preparation. This is in contrast to the conventional Hamiltonian approach where the ability to reach sufficiently low temperatures is crucial in order to access the physical properties of a many-body ground state. For synthetic systems based on ultracold atoms in optical lattices, attaining thermal equilibrium at low temperatures compared to the energy scales of the lattice potential is a key challenge, in particular for fermionic systems where the most efficient scattering channel is blocked by the Pauli principle. In this light, dissipative state preparation may be seen as an alternative to cooling and adiabatic passage mechanisms that is tailor-made to reach a specific quantum many-body state without relying on the relaxation of excitation energy by thermal exchange with a bath. Quite naturally, this approach of directly targeting a certain steady state has its own challenges, mainly relating to the complexity of the required controlled system bath interaction.\\      

The focus of this work is on inherent challenges which arise if one attempts to dissipatively prepare a particularly timely class of states, namely topological states of quantum matter \cite{HasanKane,MooreReview,XLReview}. In the context of disordered free fermionic Hamiltonians describing insulators and superconductors at mean field level, the so called periodic table of topological insulators and superconductors has been discovered \cite{classification} which systematically lists all topological states in the ten Altland Zirnbauer symmetry classes \cite{AltlandZirnbauer}. In this framework, fermionic ground states are topologically inequivalent if they cannot be continuously, i.e., without closing the energy or mobility gap of the local Hamiltonian, deformed into each other while preserving the protecting symmetries. In contrast to this theoretically well understood system of topological insulators and superconductors, the generalization of these concepts to the realm of open quantum systems out of thermal equilibrium is far from being conclusively understood despite its fundamental importance for realistic systems which are not perfectly isolated from their environment. While, in a more general context, dissipation is intuitively expected to have a mainly detrimental effect on ordering phenomena, the possibility to engineer and control dissipative processes in synthetic materials based on ultracold atoms in optical lattices enables qualitatively different scenarios. Employing controlled dissipation to {\emph{induce}} a topological state, a realization in terms of a dissipative master equation dynamics has been proposed \cite{DiehlTopDiss,BardynTopDiss} for one-dimensional topological superconductors \cite{Kitaev2001}. However, higher dimensional topological states, in particular states with a non-vanishing Chern number \cite{Chern1946,TKNN1982, Kohmoto1985}, have so far been elusive in this framework due to a fundamental competition between topology and the natural locality of the engineered system bath interaction: As we discuss in more detail below (see Section \ref{sec:locvstop}), the topology of a quantum state, in particular its Chern number, imposes fundamental constraints on how locally it can be represented in real space by, e.g., a set  of Wannier functions. More specifically, Wannier functions of a filled band with a non-vanishing Chern number can only decay algebraically in real space. This constraint in turn entail requirements regarding the complexity and non-locality of the system bath interaction that needs to be engineered in order to prepare the state of interest in a dissipative fashion.\\

In this work, we report a mechanism in momentum space coined "dissipative hole-plugging" to overcome this issue. In our present scenario, the detailed control over the system bath interaction determines how pure the dissipatively prepared steady state is rather than to which topological class it belongs. This establishes a natural notion of robustness in the dissipative preparation of topological states, where rough qualitative features of the system bath interaction determine the topology of the mixed steady state while making it pure contains some fine-tuning that may be seen as the counterpart of reaching absolute zero temperature in a Hamiltonian system. Yet, involving the interplay of two dissipative channels per degree of freedom in the target system, our construction goes conceptually beyond intuition drawn from the Hamiltonian analog. More specifically, we demonstrate that {\emph{local}} system bath engineering can result
in a master equation with a unique superfluid steady state that is characterized by a non-vanishing Chern number for fermions on a two-dimensional (2D) lattice. Quite
remarkably, in our scheme, this topology from dissipation is robust,
and exists as a stable topological phase in a non-equilibrium phase
diagram in loose analogy to a gapped phase in the Hamiltonian context. 
Furthermore, we introduce a microscopic particle number conserving model for which the dissipative analog of a mean field decoupling leads to the class of Gaussian superfluid steady states studied in our general analysis. Such a scheme can be experimentally realized with cold atoms in optical lattices combining previously reported techniques \cite{DiehlDissPrep,DiehlTopDiss,Griessner2007}, and complements recent proposals for engineering topological superfluids as Hamiltonian ground states \cite{jiang11}. We stress that the detection of the topological state in terms of its static correlations can be done as in the Hamiltonian case via state tomography \cite{alba11,kraus12,hauke14}.\\

{\emph{Outline -- }} The remainder of this article is organized as follows. In Section \ref{sec:disstop}, we set up the framework for the dissipative preparation of topological
states with a focus on challenges that arise for states with a non-vanishing Chern number. Subsequently, in Section \ref{sec:disschern}, we show how such Chern states can be dissipatively prepared in a robust way by employing the dissipative hole-plugging mechanism mentioned above. A microscopic model which, upon implementing the dissipative analog of a mean field decoupling realizes this mechanism and gives rise to a steady state with a non-vanishing Chern number is introduced and numerically analyzed in Section \ref{sec:mic}.
A concluding discussion mentioning the relation of our analysis to recent discussions on the representation of Chern insulators as tensor network states is presented in Section \ref{sec:conclusion}.

\section{Dissipative framework and topological steady states}
\label{sec:disstop}

We assume an open quantum system dynamics as described by a Lindblad master equation \cite{Lindblad1976}
\begin{align}
\dot \rho = i\left[\rho,H\right]+\sum_j\left(L_j\rho L_j^\dag -\frac{1}{2}\left\{L_j^\dag L_j,\rho\right\}\right)
\label{eqn:Lindblad}
\end{align}
for the system density matrix $\rho$ with the incoherent action of the Lindblad operators $L_j$ accounting for the coupling of the system to a bath. 
The Born-Markov approximation underlying the derivation of Eq. (\ref{eqn:Lindblad}) is well justified in a broad class of scenarios, where a system is coupled to a continuum of modes in a bath which can, e.g., be provided by phonon excitations in a Bose-Einstein condensate \cite{DiehlDissPrep}. 
We focus in the following on purely dissipative dynamics, i.e., we put $H=0$ in Eq. (\ref{eqn:Lindblad}). The physical target system that we have in mind when making this assumption consists of very weakly interacting ultracold fermionic atoms in the quasi-flat lowest Bloch band of a deep optical lattice. Under these circumstances the time evolution is dominated by the engineered dissipative dynamics, thus justifying the approximation $H=0$. While the interplay of Hamiltonian and dissipative dynamics on comparable timescales is an interesting and, at least in the quantum many-body context, largely unexplored topic in its own right, the main subject of the present work is the form of the Lindblad operators $L_j$ required to dissipatively prepare topological states of quantum matter. For our subsequent analysis, we consider lattice translation invariant Lindblad operators $L_j$ that are linear in the spinless \cite{note2} fermionic field operators $\psi_j, \psi_j^\dag$ that form a complete fermionic algebra on the 2D square lattice $\mathbb Z^2$. A master equation of this form allows solutions in the form of Gaussian states, which are characterized by their second moment correlation functions. The density matrix $\rho_k$ at lattice momentum $k$ can then be represented in the momentum space Nambu basis $(a_k,a_{-k}^\dag)$ as
\begin{align}
\rho_k=\frac{1}{2}\left(1-\vec n_k \cdot \vec \tau \right).
\label{eqn:rhodef}
\end{align} 
Here, $\tau_j$ are the Pauli matrices in Nambu space, $a_k=\frac{1}{\sqrt{N}}\sum_j\text{e}^{ikj}\psi_j$ are the Fourier transforms of the field operators, and $\lvert \vec n_k\rvert\le 1$, where $\lvert \vec n_k\rvert= 1$ holds for pure states.
In the long time limit, the density matrix $\rho$ approaches a steady state $\rho^s$. The dissipative analog of an energy gap stabilizing a Hamiltonian ground state is a so called damping gap $\kappa$ \cite{DiehlTopDiss,BardynTopDiss} (see also the subsequent Section \ref{sec:dampinggap} for a more detailed discussion), which is defined as the smallest rate at which deviations from $\rho^s$ are damped out. Formally, a damping gap implies that the zero eigenvalue of the steady state is an isolated point in the spectrum of the Liouvillian operator governing the dissipative dynamics \cite{GrafAvron}. In this sense, damping gapped steady states are the Liouvillian analogs to insulating ground states in the Hamiltonian context.

\subsection{Gaussian steady states and damping gaps}
\label{sec:dampinggap}
We now give a practical recipe for the calculation of both the steady state $\rho^s$ and the damping gap $\kappa$ of a generic Gaussian Lindblad master equation (\ref{eqn:Lindblad}).
Within the Gaussian approximation, all static information about the system density matrix $\rho$ is contained in its Gaussian correlation matrix $\Gamma$ which can be conveniently represented in the basis of the Majorana operators $c_{j,1}=\psi_j+\psi_j^\dag,~c_{j,2}=i(\psi_j^\dag-\psi_j)$ as
\begin{align}
\Gamma^{ij}_{\lambda\mu}=\frac{i}{2}\text{Tr}\left\{\rho\left[c_{i,\lambda},c_{j,\mu}\right]\right\}.
\end{align}  
With the density matrix $\rho$ evolving in time according to Eq. (\ref{eqn:Lindblad}) with $H=0$, $\Gamma$ obeys the equation of motion \cite{EisertProsen,Fleischhauer2012}
\begin{align}
\dot \Gamma =\left\{\Gamma,X\right\}-Y,
\label{eqn:Gammaeom}
\end{align}
where the matrices $X=M+M^T$ and $Y= 2i(M-M^T)$ are determined in terms of the Majorana representation of the Lindblad operators $L_j=\vec l_j^T \vec c$ via
$M=\sum_j \vec l_j \vec l_j^\dag$. Since we are dealing with lattice translation invariant systems here, it is convenient to consider the equation of motion for the Fourier transform of the correlation matrix, $\tilde \Gamma_{\lambda\mu}(k)=\frac{i}{2}\text{Tr}\left\{\rho \left[\tilde c_{k,\lambda},\tilde c_{-k,\mu}\right]\right\}$ which simply reads as
\begin{align}
\dot{\tilde \Gamma}(k)=\left\{\tilde \Gamma(k),\tilde X(k)\right\} - \tilde Y(k),
\label{eqn:eomMomentum}
\end{align}
where $\tilde X, \tilde Y$ denote the Fourier transforms of $X,Y$. Eq. (\ref{eqn:eomMomentum}) is a $2\times 2$ matrix equation for every lattice momentum $k$. 
A steady state obeys $\dot{\tilde \Gamma}_s(k)=0$. Plugging this into Eq. (\ref{eqn:eomMomentum}) yields the Sylvester equation
\begin{align}
\left\{\tilde \Gamma_s(k),\tilde X(k)\right\} = \tilde Y(k),
\label{eqn:sylvester}
\end{align}
which has a unique solution if $\tilde X(k)$ is invertible. The spectrum of the generally positive semidefinite matrix $\tilde X(k)$ determines the damping rates towards the steady state.
The minimum eigenvalue of $\tilde X(k)$ all over the Brillouin zone is called the damping gap. For a finite damping gap, the steady state $\tilde \Gamma_s(k)$ and from that, via a simple basis transformation back to the Nambu basis, the Gaussian density matrix $\rho_k^s$ itself is readily obtained by direct solution of Eq. (\ref{eqn:sylvester}).\\

\subsection{Locality versus topology in dissipative dynamics}
\label{sec:locvstop}
The major conceptual challenge in the dissipative preparation of topological states in spatial dimension $d\ge 2$ is due to the competition of topology and locality of the Lindblad operators $L_j$. Drawing intuition from Hamiltonian ground states, a generic recipe for preparing {\emph{pure}} Gaussian steady states is the following \cite{DiehlTopDiss,BardynTopDiss}: Construct a so called parent Hamiltonian $H_p=\sum_j L_j^\dag L_j$ from a complete set of anti-commuting Lindblad operators $L_j$. The ground state $\lvert G\rangle\langle G\rvert$ of this Hamiltonian is then the unique steady state of the corresponding master equation (\ref{eqn:Lindblad}) since $\lvert G\rangle$ is the only state vector annihilated by all $L_j$. In other words, the Lindblad operators are chosen as single particle operators that span the many-body ground state of $H_p$. For a lattice translation invariant parent Hamiltonian that defines a band structure, a set of Lindblad operators $L_j$ providing a real space representation of its many-body ground state corresponds to the Wannier functions of all occupied bands. However, non-trivial topological invariants characterizing the ground state impose fundamental constraints on the localization properties of the Wannier functions \cite{chernalgebraic,RashbaWannier,WannierChern,VanderbiltReview} (see Ref. \cite{JanJensEmilSebastianPeter} for a detailed recent discussion).   
The archetype of a topological invariant for band structures is the integer quantized first Chern number \cite{Chern1946,TKNN1982, Kohmoto1985}
distinguishing topologically inequivalent gapped 2D band structures. By its very definition, a non-vanishing 
Chern number implies an obstruction to finding a global smooth gauge for the associated family of Bloch functions \cite{TKNN1982, Kohmoto1985}, or, equivalently, the impossibility \cite{chernalgebraic,RashbaWannier,WannierChern} to find an exponentially localized set of Wannier functions. Hence, for the dissipative preparation of a gapped state with non-vanishing Chern number based on a parent Hamiltonian, long-ranged Lindblad operators with algebraic asymptotic decay properties would be inevitable.

\subsection{Mixed state topology}
To overcome the conflict between locality and topology discussed in the previous Section \ref{sec:locvstop}, going beyond the Hamiltonian analogy and the realm of pure steady states turns out to be crucial. For mixed states, topological properties over the lattice momentum Brillouin zone (BZ)  are well defined as long as there is a finite {\emph{purity gap}} $\lvert \vec n_k\rvert^2$ \cite{BardynTopDiss} (see Eq. (\ref{eqn:rhodef})), i.e., as long as $\rho_k$ has a finite polarization at all lattice momenta $k$. 
A rigorous and more general discussion of the notion of a purity gap can be found in Ref. \cite{JanSebastianTopDens}.
The first Chern number  \cite{Chern1946} which is the relevant topological invariant for our study is here given by 
\begin{align}
\mathcal C&= \frac{i}{2\pi}\int_{\text{BZ}}\frac{\text{Tr}\left\{ \rho_k \left[(\partial_{k_x} \rho_k),(\partial_{k_y} \rho_k)\right]\right\}}{\left(2\text{Tr}\left\{\rho_k^2\right\}-1\right)^{\frac{3}{2}}}\nonumber\\
&=\frac{1}{4\pi}\int_{\text{BZ}}\hat n_k\cdot\left[(\partial_{k_x} \hat n_k)\times(\partial_{k_y} \hat n_k)\right].
\label{eqn:Chernrho}
\end{align}
This integer quantized topological invariant measures how often the normalized polarization vector $\hat n_k =\frac{\vec n_k}{\lvert \vec n_k\rvert}$ covers the unit sphere and can only change if either the purity gap closes (rendering $\hat n_k$ ill defined at some $k$) or if the damping gap closes (giving rise to discontinuities in $k\mapsto \rho_k$). In this sense the topological steady states we are concerned with are protected by both a finite damping gap and a finite purity gap, i.e., their topology is unchanged under continuous deformations as long as both gaps are maintained. 

\begin{figure*}[htp]
\centering
\includegraphics[width=0.99\linewidth]{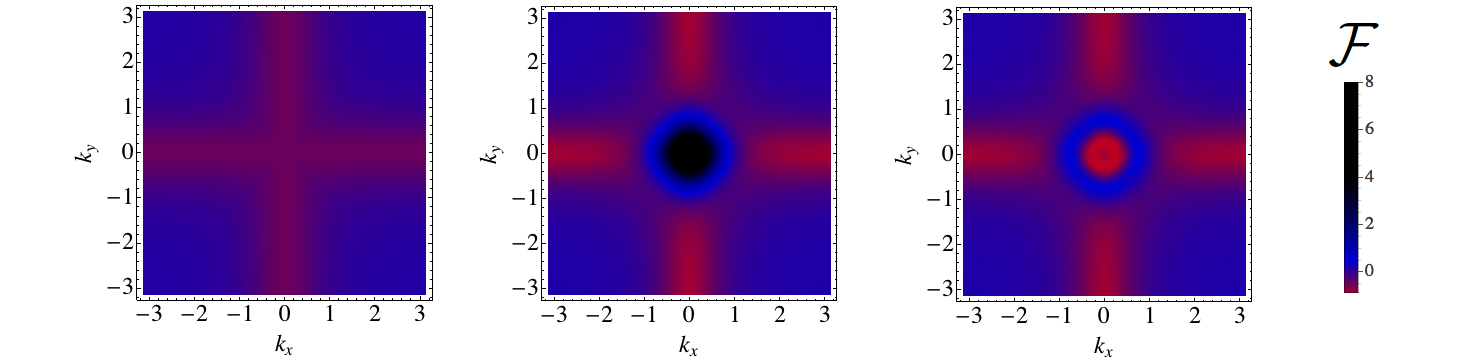}
\caption{\label{fig:Holeplugging}(Color online) Density plot of the Berry curvature $\mathcal F$ (half of integrand in Eq. (\ref{eqn:Chernrho})) as a function of momentum for $\beta =-4.0, d= g=0$ (left) $\beta =-3.0, d= g=0$ (center) and $\beta =-3.0, d=0.7, g=2.0$ (right). Performing the integral in  Eq. (\ref{eqn:Chernrho}) for the plotted function gives $\mathcal C=-1$ (left), $\mathcal C=0$ (center) and  $\mathcal C=-1$ (right), respectively. In the central plot, the peak (dark region) around $k=0$ compensates the smooth negative curvature away from the center. In the right plot the dissipative hole-plugging mechanism depletes the central peak thus maintaining the non-vanishing Chern number.
}
\end{figure*}

\section{Dissipative preparation of Chern states}
\label{sec:disschern}
To induce a steady state with non-vanishing Chern number (\ref{eqn:Chernrho}), we proceed in two steps. First, in Section \ref{sec:crit}, we construct a set $L^C$ of compactly supported Lindblad operators that yield a {\emph{critical}} Chern state as a steady state. Second, in Section \ref{sec:holeplugging} we devise a set of auxiliary Lindblad operators $L^A$ which is capable of lifting the topologically non-trivial critical point to an extended phase with a finite damping gap.\\
\subsection{Over-completeness and damping-criticality.}
\label{sec:crit}
We define Lindblad operators corresponding to a set of {\emph{non-orthonormal}} single particle states coined pseudo Wannier functions. These operators compactly supported but span only a {\emph{critical}} topological state at fine-tuned parameters. Concretely, let us consider 
\begin{align}
L_j^C= \sum_i u^C_{i-j} \psi_i + v^C_{i-j} \psi_i^\dag
\label{eqn:lcdef}
\end{align} 
where the only non-vanishing coefficients are $v^C_0=\beta,~v^C_{\pm \hat x}=v^C_{\pm \hat y}=1$ and $u^C_{\pm \hat x}= -i u^C_{\pm \hat y}=\pm 1$, supported only on nearest neighbor sites of the square lattice $\mathbb Z^2$. The steady state of the associated master equation (\ref{eqn:Lindblad}) is given by the ground state of the translation invariant parent Hamiltonian $H_p^C=\sum_j L_j^{C\,\dag} L^C_j$. The Fourier transformed Nambu spinors
\begin{align}
&B_k^C\equiv(\tilde u^C_k,\tilde v^C_k)^T=\nonumber\\
&\left(2i\left(\sin(k_x)+i\sin(k_y)\right),\beta+2(\cos(k_x)+\cos(k_y))\right)^T
\label{eqn:pseudobloch}
\end{align}
are coined pseudo Bloch functions and
are non-vanishing all over the BZ except at isolated values of $\beta$. At $\beta=-4$, for example, $(\tilde u^C_0, \tilde v^C_0)=(0,0)$. Such a zero gives rise to a closing of the damping gap $\kappa^C_k=\lvert \tilde u^C_k\rvert^2+\vert \tilde v^C_k \rvert^2$, i.e., a critical damping behavior. Yet, the normalized polarisation vector $\hat n_k$ of the steady state $\rho_k^s$ continuously approaches the value $\hat n_0=\hat e_3$ as $k\rightarrow 0$ and can thus be defined all over the BZ even at the critical point $\beta=-4$. Direct calculation of the associated Chern number using Eq. (\ref{eqn:Chernrho}) yields $\mathcal C=-1$. However, $\hat n_k$ is not real analytic in $k=0$ but only finitely differentiable. As a consequence $\rho_k^s$ at $\beta=-4$ corresponds to a {\emph{critical}} Chern-state that can have algebraically decaying correlations. The pseudo Wannier operators (\ref{eqn:lcdef}) form an {\emph{overcomplete}} set of that critical state since the vanishing pseudo Bloch function $B_0^C$ (see Eq. (\ref{eqn:pseudobloch})) precludes their linear independence. Away from such isolated parameter-points, the pseudo Bloch functions are non-vanishing all over the BZ and $\mathcal C=0$.
This phenomenology is generic and not due to an unfortunate choice of our model: As has been extensively analyzed for free electrons in a perpendicular magnetic field, localized pseudo Wannier functions spanning the lowest Landau level (LLL) are necessarily overcomplete due to the non-zero Chern number of the LLL (see, e.g., Ref. \cite{RashbaWannier}). For the same topological reason, the above pseudo Bloch functions $B_k^C$ must exhibit a zero associated with a critical damping behavior at the isolated topologically non-trivial points.  A dissipative preparation of such a critical state involves fine tuning since an infinitesimal deviation removes the essential zero from the pseudo Bloch functions, in turn rendering the state topologically trivial. We will now devise an auxiliary set of Lindblad operators $L_j^A$ which is capable of lifting the critical state resulting only from the fine-tuned $L_j^C$ operators to a gapped extended phase.\\

\subsection{Dissipative hole-plugging mechanism}
\label{sec:holeplugging}
\begin{figure}[htp]
\centering
\includegraphics[width=0.49\columnwidth]{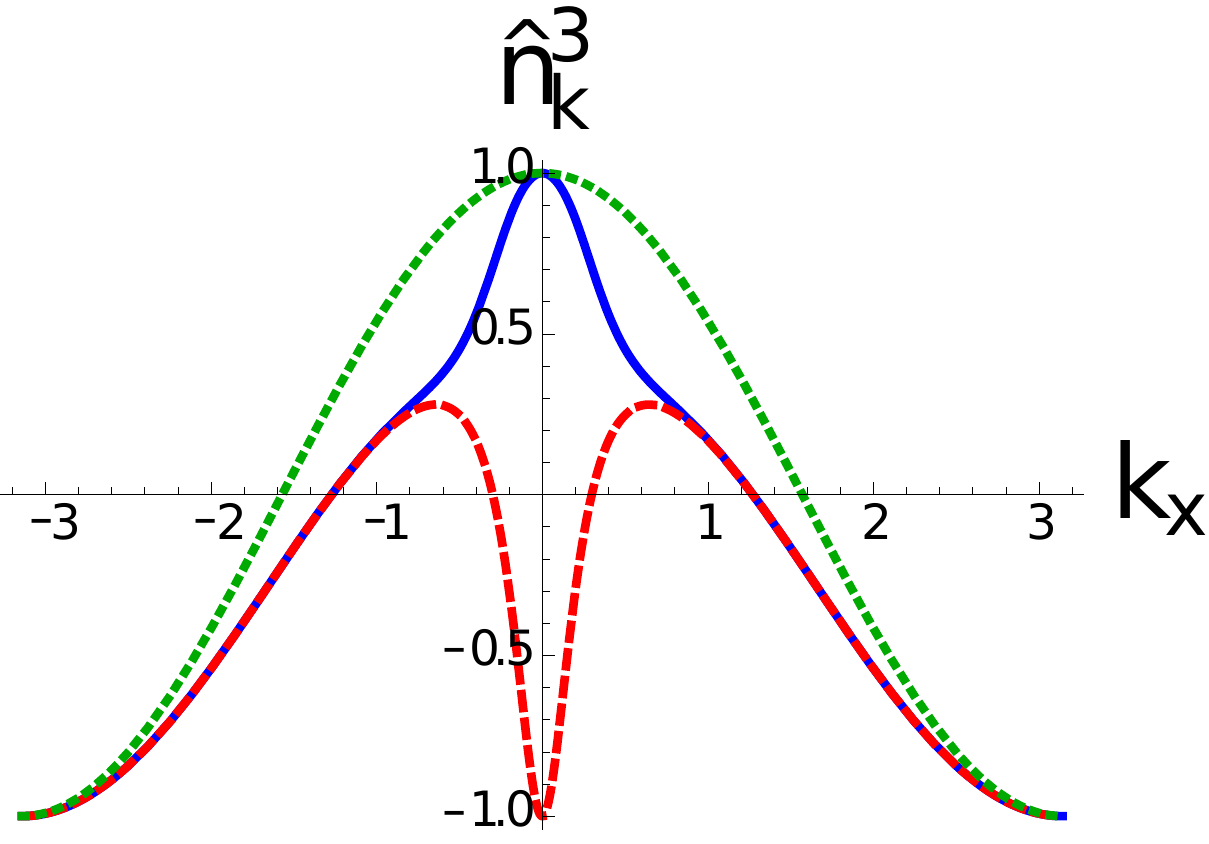}
\includegraphics[width=0.49\columnwidth]{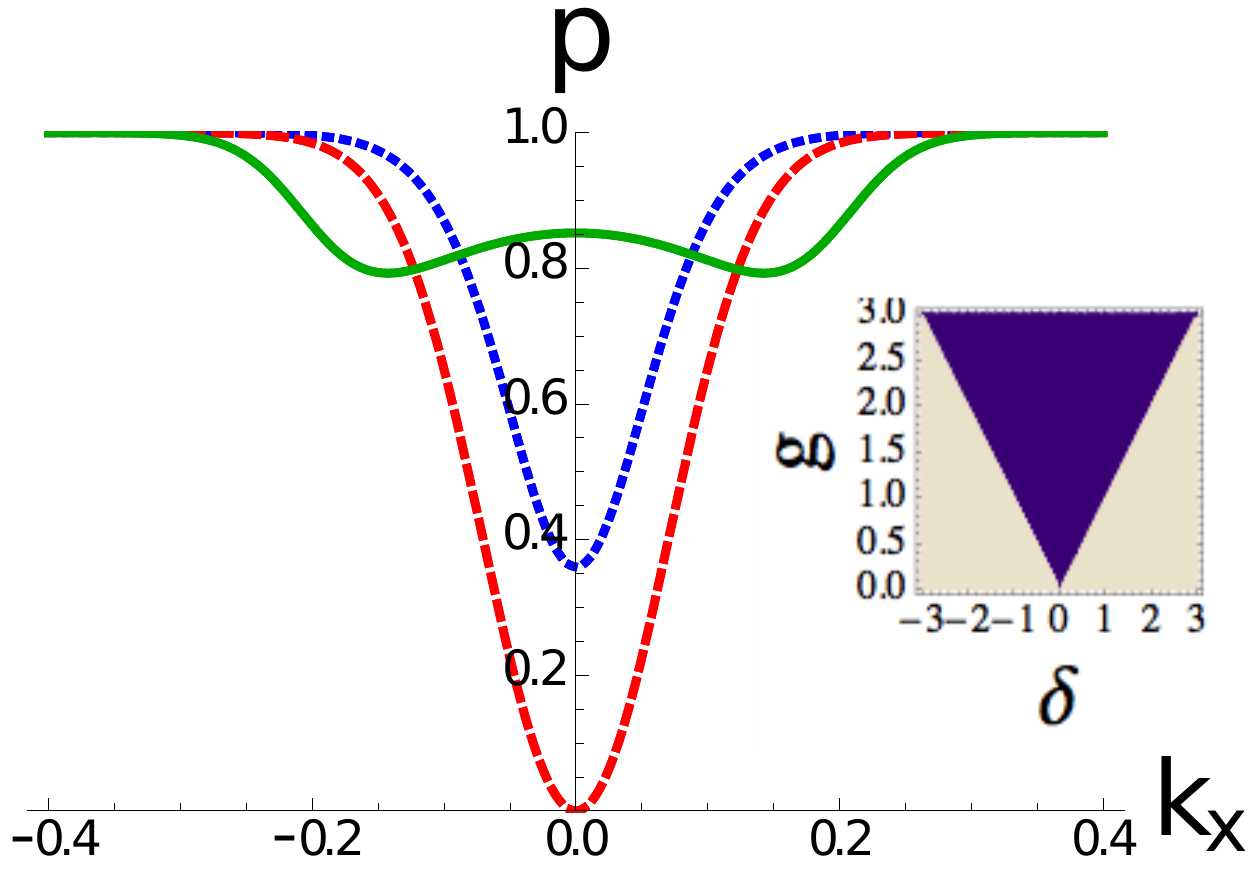}
\caption{\label{fig:puritygapclosing}(color online) Left panel: $\hat n_k^3$ as a function of $k_x$ at $k_y=0$ for $\delta=0,d=g=0$ (green dotted), for $\delta=0.5,d=g=0$ (Red dashed), and for $\delta=0.5,d=0.7,g=1.0$ (blue solid). Right panel: Purity gap $p=\lvert\vec n_k\rvert^2$ of the steady state $\rho_k^s$ as a function of $k_x$ at $k_y=0, ~\delta=d=0.2$. Gap at $g=0.1$ (blue dotted) in the topologically trivial phase, purity critical point at $g=0.2$ (red dashed), gap at $g=1.0$ (green solid) in the non-trivial phase. Inset: Phase diagram of the steady state as a function of $\delta=-4-\beta$ and $g$. $d=1.0$ is fixed. The purple region has Chern number $\mathcal C=-1$, while $\mathcal C=0$ in the bright region. The purity gap closes at the transition lines. The damping gap is finite everywhere except at the critical point $\delta=g=0$.}
\end{figure}

Deviating by $\delta$ from the topologically non-trivial critical point, i.e., $\beta=-4-\delta$ in (\ref{eqn:lcdef}) may be viewed as tearing a hole into the smooth winding of $\hat n_k$ as a function of $k$. The simplest way to see this is to consider the third component $\hat n_k^3$ of $\hat n_k$ in the steady state. With the Lindblad operators given by Eq. (\ref{eqn:lcdef}), we have $\hat n_k^3=\lvert \tilde u_k^C\vert^2-\lvert \tilde v_k^C\vert^2$. Indeed, a finite value of $\tilde v^C_0=\delta$ enforces that $\hat n_0=-\hat e_3$ since $\tilde u_0^C=0$ (see Eq. (\ref{eqn:pseudobloch})). In contrast, at small $k>\delta$, $\hat n \approx \hat e^3$ since $\tilde u_k^C=2i(k_x+ik_y)+\mathcal O(k^2)$. This rapid change in $\hat n_k^3$ compensates the almost complete smooth winding over the rest of the BZ (compare Fig. \ref{fig:Holeplugging} left and central panel or the red dashed and green dotted left plots in Fig. \ref{fig:puritygapclosing}) rendering the state topologically trivial even at infinitesimal $\delta$. From a more physical perspective, $(1-\hat n_k^3)/2$ measures the occupation number $\langle a_k^\dag a_k\rangle$ which is rapidly changing around $k=0$ for finite $\delta$ (see red dashed plot in the left panel of Fig. \ref{fig:puritygapclosing}).

To compensate for this "topological leak", we propose a {\emph{dissipative hole-plugging mechanism}} which stabilizes the smooth winding of $\hat n_k$ even at finite deviations from the critical point. More specifically, we introduce the auxiliary Lindblad operators $\tilde L_k^A=\tilde u_k^A a_k$ in momentum space which have only an annihilation part ($\tilde v_k^A=0$) with, e.g., a Gaussian weight function $\tilde u_k^A=g \text{e}^{-k^2/d^2}$. These operators act selectively in momentum space: they prevent the unwanted occupation of the $k=0$ mode as long as $g>\delta$, but their action becomes irrelevant for $k\gg d$. The Gaussian weight function in the definition of $\tilde L_k^A$ makes the favorable exponential localization properties manifest also in real space, alternative choices  are, however, readily conceivable. The key qualitative point of the dissipative hole-plugging mechanism is the momentum selective depletion around $k=0$ which can also be achieved with a lattice regularized weight function and a finite number of Fourier modes, i.e., with Lindblad operators that are compactly supported in real space. Our dissipative hole-plugging mechanism is illustrated in terms of the Berry curvature $\mathcal F=\frac{1}{2}\,\hat n_k\cdot\left[(\partial_{k_x} \hat n_k)\times(\partial_{k_y} \hat n_k)\right]$, i.e., the integrand of Eq. (\ref{eqn:Chernrho}) in Fig. \ref{fig:Holeplugging}. In the left panel, $\mathcal F$ is shown exactly at the topological critical point $\beta=-4$ where $\mathcal C=-1$. In the central panel, $\mathcal F$ is plotted somewhat away from the critical point at $\beta=-3$.  The peak (dark region) around $k=0$ compensates the negative curvature away from $k=0$ thus giving rise to $\mathcal C=0$. In the right panel this dark center region is suppressed by the action of the $L^A_j$ jump operators thus maintaining $\mathcal C=-1$ even away from the critical point.

We now demonstrate that our hole plugging mechanism leads to a topologically non-trivial steady state in a finite parameter range around the critical point $\beta=-4$. We calculate the Chern number of the steady state obtained from the interplay of the Lindblad operators $L^C_j$ and $L^A_j$ while monitoring both its damping gap $\kappa_k$ \cite{BardynTopDiss} and its purity gap $\lvert \hat n_k\rvert^2$. We find that the auxiliary jump operators $L^A_j$ are capable of lifting the isolated points at which the $L^C_j$ become topologically non-trivial to an extended phase (see inset in right panel of Fig. \ref{fig:puritygapclosing}). 
If we start in the absence of $L^A_j$ with a topologically trivial $\beta = -4-\delta$, i.e., detuned from the critical point by $\delta>0$ and switch on $L^A_j$ by ramping up $g$, we observe a topological transition associated with a purity gap closing at $g=\delta$ (see Fig. \ref{fig:puritygapclosing} right panel). At $g>\delta$, the purity gap reopens and the steady state has Chern number $\mathcal C=-1$. The damping gap stays finite throughout this procedure. 

We stress that the topological properties of the steady state are quite insensitive to the value of $d$. However, a larger value of $d$ leads to a larger region around $k=0$ where the purity gap differs significantly from one, i.e., where the steady state is significantly mixed. This reflects the earlier mentioned tradeoff between locality of the $L^A_j$ operators and the purity of the dissipatively prepared Chern state. Another purity gap closing which would render the Chern number (\ref{eqn:Chernrho}) un-defined does, however, not occur in a wide parameter range, even if the Gaussian weight function $g \text{e}^{-k^2/d^2}$ is approximated by a few Fourier modes.\\

\section{Microscopic implementation}
\label{sec:mic}
 
So far, we have generally analyzed how a superfluid steady state with a non-vanishing Chern number can occur at the level of a Gaussian Lindblad master equation (\ref{eqn:Lindblad}). In the Hamiltonian case, a superconducting condensate  arises at mean field level in the thermodynamic limit from an interacting particle number conserving microscopic Hamiltonian. Also in our present dissipative framework, an effective quadratic master equation with spontaneously broken $U(1)$ symmetry arises from a microscopic, particle number conserving model described by a master equation that is {\emph{quartic} in the field operators. In the following, we introduce such a model and outline how it can be experimentally implemented with cold atoms in optical lattices. As we  confirm numerically, the phenomenology described above, in particular our {\emph{dissipative hole-plugging}} mechanism is obtained in a mean field approximation analogous to the one introduced in Ref.~\cite{DiehlTopDiss}.

\subsection{Particle number conserving quartic model}
Our model again consists of a near-critical and -topological set of Lindblad operators $\ell^C$, and an auxiliary set $\ell^A$. The Lindblad operators have the number conserving bilinear form $ \ell^\alpha_i = C^{\alpha\, \dag}_i A^\alpha_i,~\alpha=C,A$ with the creation $C^{\alpha\,\dag}_i = \sum_jv_{j-i}^\alpha\psi_j^\dag$ and annihilation parts $A^{\alpha\,\dag}_i = \sum_j u_{j-i}^\alpha\psi_j$. For $\alpha=C$, the coefficients $u_j^C,v_j^C$ are the same as in Eq.~\eqref{eqn:lcdef}. The experimental implementation of Lindblad operators of such a form has been discussed in Ref. \cite{BardynTopDiss}. The auxiliary operators $\ell^A_j$ are chosen such that particles are pumped out of the central region of the Brillouin zone into the higher momentum states thus reflecting the depletion of low momenta which is at the heart of our hole plugging mechanism. For atoms in optical lattices this can be achieved by momentum selective pumping techniques as described in \cite{Griessner2007}. In momentum space, $\tilde \ell^A_k  = \sum_q \tilde C^{A\,\dag}_{q-k}  \tilde A^{A}_q$, with $\tilde C^{A\,\dag}_k = \tilde v^A_k  a^\dag_k , \tilde A^A_k  = \tilde u^A_k  a_k$. The momentum selective functions are ideally of the form $\tilde u^A_k = g_u e^{-k^2/ d_u^2}$ removing particles from the central region, and $\tilde v^A_k = g_v \sum_i e^{-(k -\pi_i)^2/d_v^2}$ describing their reappearance at high momenta $\pi_i \in \{(0,\pi), (\pi,0),(\pi,\pi)\}$. The key qualitative point to the form of $\ell^A_k$ that has to be reflected in an experimental realization is the dominance of processes taking a particle at the center of the Brillouin zone and transferring a momentum of order $\pi$.

\subsection{Dissipative mean field theory}
A self-consistent mean field theory can be derived building up on Ref. \cite{BardynTopDiss}. We start from the master equation 
\begin{eqnarray}
\partial_t \rho = \mathcal L^C[\rho] + \mathcal L^A[\rho],
\end{eqnarray}
where $\mathcal L^\alpha[\rho]=\sum_j \left(\ell^\alpha_j \rho \ell^{\alpha\,\dag}_j-\frac{1}{2}\left\{\ell^{\alpha\,\dag}_j\ell^\alpha_j,\rho\right\}\right),\quad \alpha=C,A$. We write out this quartic master equation in momentum space, and make the ansatz $\rho= \prod_ {k}'\rho_ {k}$, where $\rho_ {k}$ describes the mode pair $\{ {k},- {k}\}$ and obeys $\mathrm{Tr}_ {k}\rho_ {k} = 1$. $ \prod_ {k}'$ reminds that the product is taken over half of the Brillouin zone only, e.g. the upper half. We then focus on one particular momentum mode pair $\{ {p},- {p}\}$, and keep only terms which are quadratic in operators associated to this mode pair. By means of the prescription $\rho_ {p} = \mathrm{Tr}_{\neq  {p}} \mathcal L [\rho]$,  we obtain an evolution equation for $\rho_ {p}$ with coefficients $C_i^\alpha$, which are governed by the mean fields of the remaining modes in the system,
\begin{eqnarray}\label{eq:eqC}
\partial_t \rho_ {p} &= &\sum_{\alpha = C,A} \Big\{ C_1^\alpha |\tilde v^\alpha_ {p}|^2 \mathcal L_{a_ {p}^\dag,a_ {p}}[\rho_ {p}] 
  + C_2^\alpha | \tilde u^\alpha_ {p}|^2  \mathcal L_{a_ {p},a_ {p}^\dag}[\rho_ {p}]   \nonumber\\
& & - C_3^\alpha  \tilde v^\alpha_ {p} \tilde u^{\alpha\, *}_{- {p}}\mathcal L_{a_ {p}^\dag,a_{- {p}}^\dag}[\rho_ {p}] 
  - C_3^{\alpha \,*} \tilde u^\alpha_{- {p}} \tilde v^{\alpha\,*}_{ {p}}\mathcal L_{a_{- {p}},a_ {p}}[\rho_ {p}]  \nonumber\\
&&   +  \{ {p}\to - {p}\} \Big\}, \label{eq:leff}
\end{eqnarray}
with abbreviation $\mathcal L_{a,b}[\rho] =  a\rho b - \tfrac{1}{2} \{ba,\rho\}$ and 
\begin{eqnarray}
C_1^\alpha &=& \sum_{ {q}\neq  {p} } | \tilde u^\alpha_ {q} |^2\langle a^\dag_ {q}  a_ {q}\rangle, \,\,
C_2^\alpha = \sum_{ {q}\neq  {p} } |\tilde v^\alpha_ {q}|^2 (1 - \langle  a^\dag_ {q} a_ {q}  \rangle),\nonumber\\
C_3^\alpha &=& \sum_{ {q}\neq  {p} } \tilde u^\alpha_ {q} \tilde v^{\alpha\,*}_{- {q}} \langle a_{- {q}}  a_ {q}\rangle;
\end{eqnarray}
we note $ \langle a_{- {q}}  a_{ {q}}\rangle^* = \langle a^\dag_ {q}  a_{- {q}}^\dag\rangle $, and that the constraint on the sum can be neglected in the thermodynamic limit.

In the absence of $\mathcal L^A$, the stationary state is known explicity, using the equivalence of fixed number and fixed phase wavefunctions in the thermodynamic limit, and the exact knowledge of the fixed number state annihilated by the set $\ell^C_i$ \cite{BardynTopDiss}. It is given by the pure density matrix $\rho_D = |\psi\rangle \langle \psi|$, where $ |\psi\rangle =  \prod'_{ {q}} \mathcal N_ {q} (1 + \frac{\tilde v^C_ {q}}{\tilde u^C_ {q}}  a_{- {q}}^\dag a_ {q}^\dag) | 0\rangle$ and $\mathcal N_ {q} = 1/\sqrt{1 + |\tilde v^C_ {q}/\tilde u^C_ {q}|^2}$. In this case, the averages can be evaluated explicitly, $\langle a^\dag_ {q}  a_ {q}\rangle = |\bar v^C_ {q}|^2,  \langle a_ {q}  a^\dag_ {q}\rangle = |\bar u^C_ {q}|^2, \langle a_{- {q}}  a_ {q}\rangle = \bar v^C_{- {q}} \bar u^{C\,*}_ {q} , \quad \langle a^\dag_ {q}  a_{- {q}}^\dag\rangle =  \bar v^{C\,*}_{- {q}} \bar  u^C_ {q}$, where $ \bar v^C_ {q} = \tilde v^C_ {q} /\sqrt{|\tilde u^C_ {q}|^2+ |\tilde v^C_ {q}|^2}$ and analogous for $ \bar u^C_ {q}$. Note that, in this case, one common real number $C_0 = C_1^C = C_2^C =C_3^C =C_3^{C\,*} =  \sum_{ {q}\neq  {p} }\frac{ | \tilde u^C_ {q}  \tilde v^C_ {q} |^2}{|\tilde u^C_ {q}|^2+ |\tilde v^C_ {q}|^2}$ can be factored out of Eq.~\eqref{eq:leff}. The resulting linearized Lindblad operators coincide with those defined in Eq. (\ref{eqn:lcdef}) in the main text (at half filling, and up to an irrelevant relative phase which reflects spontaneous symmetry breaking).  In other words, the product of the creation and annhilation part in $\ell_i^C$ can be linearized and transforms into its sum \cite{DiehlTopDiss}, thus yielding precisely the form displayed in Eq.~(\ref{eqn:lcdef}) at half filling.  This solution serves as a starting point to evaluate the stationary state of the master equation with both sets of Lindblad operators $\ell^C, \ell^A$ at mean field level. 

When $\mathcal L^A$ is added, the stationary state is no longer known explicitly. However, a self-consistent mean field theory can still be constructed. To this end, we derive the evolution of the covariances for the mode pair $\{ {p},- {p}\}$,
\begin{align}
&\partial_t \left(\begin{array}{c}
\langle a^\dag_ {p} a_ {p} \rangle \\
\langle a_{- {p}} a_{ {p}} \rangle \\
\langle a^\dag_{ {p}} a^\dag_{- {p}} \rangle 
\end{array}\right) = \nonumber\\
 &\left(\begin{array}{ccc}
 - \kappa_ {p} & \nu^*_ {p} & \nu_ {p}  \\
0 &- \kappa_ {p} & 0 \\
0 & 0 & -\kappa_ {p}
\end{array}\right)
 \left(\begin{array}{c}
\langle a^\dag_ {p} a_ {p} \rangle \\
\langle a_{- {p}} a_{ {p}} \rangle \\
\langle a^\dag_{ {p}} a^\dag_{- {p}} \rangle 
\end{array}\right)
+  \left(\begin{array}{c}
\mu_ {p} \\
\lambda_ {p} \\
\lambda^*_ {p}
\end{array}\right),\nonumber\\
\end{align}
where
\begin{eqnarray}
 \kappa_ {p} &=& \sum_\alpha (C_1^\alpha |\tilde v^\alpha_ {p}|^2 + C_2^\alpha |\tilde u^\alpha_ {p}|^2 ),\\\nonumber
 \nu_ {p} &=& \tfrac{1}{2} \sum_\alpha C_3^\alpha (\tilde  v^\alpha_{- {p}} \tilde u^{\alpha\, *}_ {p} + \tilde v^\alpha_ {p}   \tilde u^{\alpha\, *}_{- {p}}  ),\\\nonumber
\mu_ {p} &=&  \sum_\alpha C_1^\alpha |\tilde v^\alpha_ {p}|^2  , \,
\lambda_ {p} =  \tfrac{1}{2} \sum_\alpha C_3^\alpha (\tilde v^\alpha_{- {p}} \tilde u^{\alpha\, *}_ {p}  - \tilde v^\alpha_ {p}   \tilde u^{\alpha\, *}_{- {p}}   ).
\end{eqnarray}
Here, we have used the property $|\tilde v^\alpha_ {p}|^2 = |\tilde v^\alpha_{- {p}}|^2$ and analogous for $\tilde u_ {p}$, exhibited by our Lindblad operators. 
The coefficients $C_i^\alpha$ make these equations non-linear. The implicit equations for the stationary state read as 
\begin{eqnarray}\label{eq:nonlinsol}
\left(\begin{array}{c}
\langle a^\dag_ {p} a_ {p} \rangle \\
\langle a_{- {p}} a_{ {p}} \rangle \\
\langle a^\dag_{ {p}} a^\dag_{- {p}} \rangle 
\end{array}\right) = 
 \frac{1}{\kappa_ {p}} \left(\begin{array}{ccc}
1 &  \frac{\nu^*_ {p} }{\kappa_ {p}} & \frac{\nu_ {p}}{\kappa_ {p}}  \\
0 &1& 0 \\
0 & 0 & 1
\end{array}\right)  \left(\begin{array}{c}
\mu_ {p} \\
\lambda_ {p} \\
\lambda^*_ {p}
\end{array}\right).\nonumber\\
\end{eqnarray}
These equations can be solved iteratively, starting from the known solution of $\mathcal L^C$ alone. Qualitative properties of the solution can be discussed on the basis of the localization properties of the functions $\tilde u^A_ {q}, \tilde v^A_ {q}$ in momentum space. In particular, based on Eq.~\eqref{eq:eqC}, we expect modifications of the solution for $\mathcal L^C$ only in the central and edge regions of the Brillouin zone, as clearly effective annhilation (creation) of particles takes place in the center (edges) of the Brillouin zone. In addition, the off-diagonal contributions to the effective non-topological Liouvillian are exponentially small.

\begin{figure}[htp]
\centering
\includegraphics[width=0.6\columnwidth]{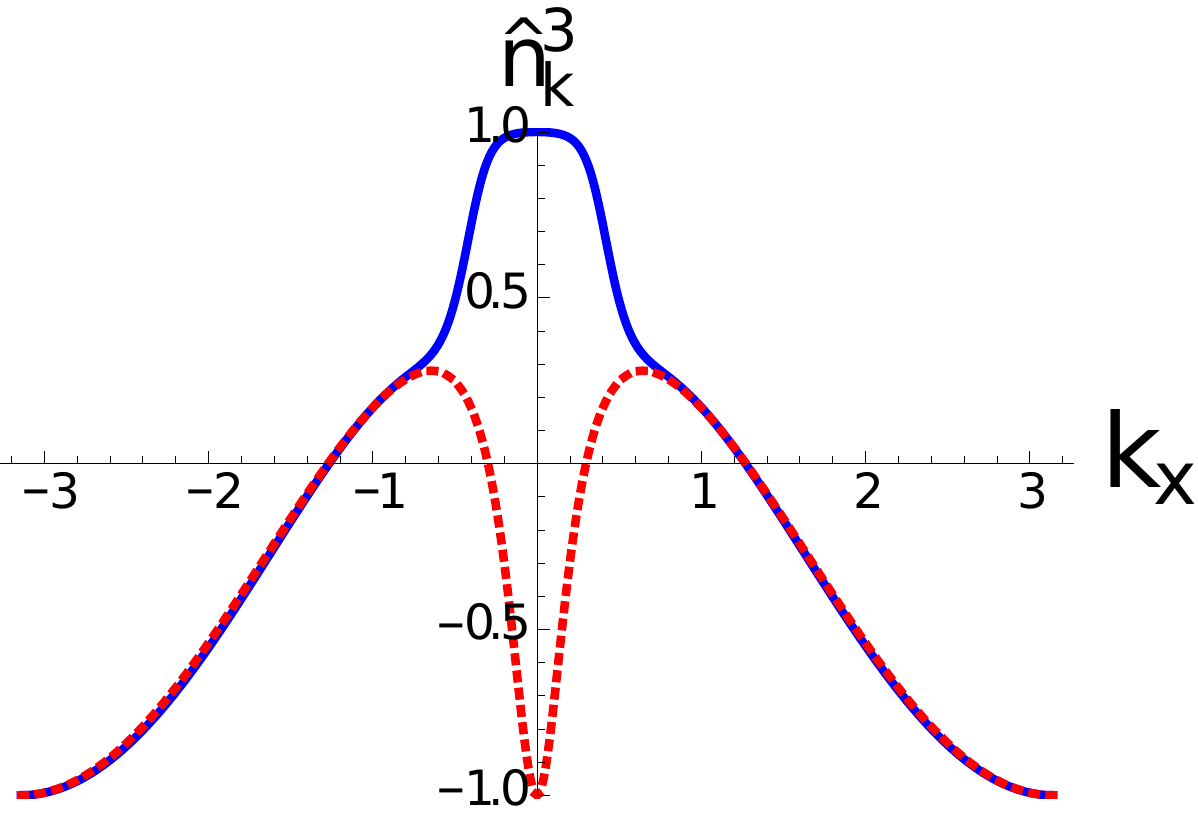}
\caption{\label{fig:microplugging} $\hat n^3_k=-\text{Tr}\left\{\rho_k^s\tau_3\right\}$ as a function of $k_x$ at $k_y=0$. Red dashed plot for $\delta=-4-\beta = 0.5$ in the presence of $\ell^C_j$ only. Blue solid plot shows the self-consistent solution of Eq. (\ref{eq:nonlinsol}) for $\delta=0.5, g_u=g_v=5.0,d_u=0.5,d_v=1.5$ in the presence of both $\ell^C_j$ and $\ell^A_j$ on a lattice of $501\times 501$ sites.}
\end{figure}

\subsection{Numerical results}
Our numerical simulations are done for a lattice of $501\times 501$ sites with periodic boundary conditions.
The above general picture, in particular the efficiency of the dissipative hole-plugging mechanism to stabilize a state with non-vanishing Chern number, is fully confirmed by the numerical analysis of this microscopic model. In Fig. \ref{fig:microplugging}, we give an explicit example demonstrating how the self consistent solution of Eq. (\ref{eq:nonlinsol}) is capable of achieving the dissipative  hole-plugging mechanism.\\

\section{Concluding discussion} 
\label{sec:conclusion}
  
The target state of the explicit construction presented here resembles the $p+ip$ superconducting ground state introduced by Read and Green \cite{ReadGreen}, i.e., a topological state with non-vanishing Chern number in symmetry class D \cite{AltlandZirnbauer}. The generalization to gapped quantum anomalous Hall states aka Chern insulators \cite{QAH} in symmetry class A, however, is straightforward. Our present analysis focuses on the possibility of preparing gapped states with a non-vanishing Chern number by means of short-ranged Lindblad operators at the expense of obtaining a mixed steady state. Alternatively, the implementation of long-ranged, i.e., algebraically decaying Lindblad operators leading to a pure Chern state as a steady state might be an interesting future direction.  

The criticality of the steady state supported by the $L^C$ operators (see Eq. (\ref{eqn:lcdef})) is in some analogy to Ref. \cite{DubailReadCriticalChern}, where a tensor network state representing a critical Chern state is constructed. While it may be impossible to represent pure non-critical states with non-vanishing Chern number as tensor network states, resorting to mixed states that are uniquely associated with a pure state by replacing $\vec n_k$ with $\hat n_k$ in Eq. (\ref{eqn:rhodef}) might be fruitful for systematic approximations of such scenarios, where the deviation from the gapped pure state in the expectation value of any observable can be bounded in terms of the purity gap.

As mentioned above, the Chern number of a mixed state as defined in Eq. (\ref{eqn:Chernrho}) is an observable property of the density matrix. However, the general relation between natural response quantities and topological invariants in open quantum systems is still an open question that is subject of ongoing research.\\

\section*{Acknowledgements}   
{We acknowledge interesting discussions with Emil J. Bergholtz, Jens Eisert, and Tao Shi on related projects. Support from the ERC \je{Synergy Grant} UQUAM, the START Grant No. Y 581-N16, the SFB FoQuS (FWF Project No. F4006- N16), and the German Research Foundation (DPG) through ZUK 64 is gratefully acknowledged.}

\bibliographystyle{apsrev}

\begin{thebibliography}{7}

\bibitem{poyatos96}
J. F. Poyatos, J. I. Cirac, J. I. and P. Zoller, Phys. Rev. Lett. {\bf 77} 4728 (1996).


\bibitem{DiehlDissPrep}
S.~Diehl, A.~Micheli, A.~Kantian, B.~Kraus, H.~ P.~ B\"uchler, and P.~Zoller,
Nature Physics {\bf{4}}, 878 (2008).

\bibitem{Verstraete2008}
F.\ Verstraete, M.\ M.\ Wolf, J.\ I.\ Cirac,
Nature Physics {\bf{5}}, 633 (2009).

\bibitem{Weimer2010}
H.\ Weimer, M.\ M\"uller, I.\ Lesanovsky, P.\ Zoller, H.\ P.\ B\"uchler, Nature Phys. {\bf{6}}, 382 (2010). 

\bibitem{barreiroNature}
   J.\ Barreiro, M.\ M\"uller, P.\ Schindler, D.\ Nigg, T.\ Monz, M.\ Chwalla, M.\ Hennrich, C.\  F.\ Roos, P.\ Zoller, R.\ Blatt, Nature {\bf{470}}, 486 (2011).

\bibitem{Krauter2011}
H.\ Krauter, C.\ A.\ Muschik, K.\ Jensen, W.\ Wasilewski, J.\ M.\ Petersen, J.\ I.\ Cirac, E.\ S.\ Polzik, Phys.\ Rev.\ Lett. {\bf{107}}, 080503 (2011).

\bibitem{kapit14}
E. Kapit, M. Hafezi and S. H. Simon, Phys. Rev. X {\bf 4} 031039 (2014).


\bibitem{ZwergerReview}
I. Bloch, J. Dalibard, and W. Zwerger, Rev.\ Mod.\ Phys.\  {\bf 80}, 885 (2008).


\bibitem{LewensteinReview}
M.\ Lewenstein, A.\ Sanpera, V.\ Ahufinger, B.\ Damski, A.\ Sen, and  U.\ Sen, Advances in Physics {\bf 56(2)}, 243 (2007). 


\bibitem{ColdFermionReview}
S. Giorgini, L. P. Pitaevskii, S. Stringari, Theory of ultracold atomic Fermi gases, Rev. Mod. Phys. {\bf{80}}, 1215 (2008).



  \bibitem{jotzu14}
G. Jotzu et al., Nature {\bf  515} 237-240 (2014).

 \bibitem{aidelsburger14}
M. Aidelsburger et al., {arXiv:1407.4205} (2014).


\bibitem{HasanKane}
M.\ Z.\ Hasan and C.\ L.\ Kane, Rev.\ Mod.\ Phys.\  {\bf 82}, 3045 (2010).
	
\bibitem{MooreReview}
J.\ E.\ Moore, Nature {\bf 464}, 194 (2010).

\bibitem{XLReview}
 X.-L.\ Qi and S.-C.\ Zhang, Rev.\ Mod.\ Phys.\ {\bf 83}, 1057 (2011).



\bibitem{classification} 
	\je{A.\ P.\ Schnyder, S.\ Ryu, A.\ Furusaki, and A.\ W.\ W.\ Ludwig, Phys.\ Rev.\ B {\bf 78}, 195125 (2008);
	A.\ Kitaev, AIP Conference Proceedings {\bf 1134}, 22 (2009);
	S.\ Ryu, A.\ P.\ Schnyder, A.\ Furusaki, and A.\ W.\ W.\ Ludwig, New J.\ Phys.\ {\bf 12}, 065010 (2010).}
	
\bibitem{AltlandZirnbauer}
	{A. Altland and M.\ R.\ Zirnbauer, Phys.\ Rev.\ B {\bf 55}, 1142 (1997).}	
	

\bibitem{DiehlTopDiss}
S.~Diehl, E.~Rico, M.~A. Baranov, and P.~Zoller,
Nature Physics {\bf{7}}, 971 (2011).



\bibitem{BardynTopDiss} 
	C.-E.\ Bardyn, M.\ A.\ Baranov, C.\ V.\ Kraus, E.\ Rico, A.\ Imamoglu, P.\ Zoller, S.\ Diehl, New J. Phys. {\bf 15}, 085001  (2013).

\bibitem{Kitaev2001}
	A.\ Kitaev,
  	Physics-Uspekhi {\bf 44}, 131 (2001).




\bibitem{Chern1946}
S.\ S.\ Chern, The Annals of Mathematics, {\bf 47} (1), 85 (1946).

\bibitem{TKNN1982}
D.\ J.\ Thouless, M.\ Kohmoto, M.\ P.\ Nightingale, and M.\ den
Nijs, Phys.\ Rev.\ Lett.\ {\bf 49}, 405 (1982).

\bibitem[{ {Kohmoto}(1985)}]{Kohmoto1985}
\bibinfo{author}{ {M.}~ {Kohmoto}},
  \bibinfo{journal}{Annals of Physics} \textbf{\bibinfo{volume}{160}},
  \bibinfo{pages}{343} (\bibinfo{year}{1985}).
 
\bibitem{Griessner2007}
A.\ Griessner, A.\ J.\ Daley, S.\ R.\ Clark, D.\ Jaksch, P.\ Zoller, New J. Phys {\emph 9}, 44 (2007).  
  
  \bibitem{jiang11}
L. Jiang et al., Phys. Rev. Lett. {\bf 106} 220402 (2011).

\bibitem{alba11}
E. Alba et al., Phys. Rev. Lett. {\bf 107} 235301(2011).

\bibitem{kraus12}
C. V. Kraus et al., New J. Phys. {\bf 14}, 113036 (2012).


\bibitem{hauke14}
P. Hauke, M. Lewenstein, A. Eckardt, Phys. Rev. Lett. {\bf 113} 045303(2014).



\bibitem{Lindblad1976}
G. Lindblad, Commun. Math. Phys. {\bf 48} (2), 119 (1976).


\bibitem{note2}
Our analysis can be generalized to consider spin and other internal degrees of freedom in a straight forward way. For the sake of simplicity, we here describe the spinless case which is sufficient for our subsequent construction.

\bibitem{GrafAvron}
J.\ E.\ Avron, M.\ Fraas, G.\ M.\ Graf, Journal of Statistical Physics {\bf{148 (5)}}, 800 (2012). 
J.\ E.\ Avron, M.\ Fraas, G.\ M.\ Graf, O.\ Kenneth, New J. Phys. {\bf{13}}, 053042 (2011).


\bibitem{EisertProsen}
J.\ Eisert, T.\ Prosen, arXiv:1012.5013 (2010).

\bibitem{Fleischhauer2012}
M.\ Hoening, M.\ Moos, M.\ Fleischhauer, Phys. Rev. A {\bf{86}}, 013606 (2012). 

\bibitem{chernalgebraic} 
	D.\ J.\ Thouless, J.\ Phys.\ C {\bf 17},  L325 (1984).
	%Wannier functions for magnetic sub-bands,

\bibitem{RashbaWannier} 
E.\ I.\ Rashba, L.\ E.\ Zhukov, and A.\ L.\ Efros,
Phys.\ Rev.\ B 55, 5306 (1997). 

\bibitem{WannierChern}
	{C.\ Brouder, G.\ Panati, M.\ Calandra, C.\ Mourougane, and N.\ Marzari,
	Phys.\ Rev.\ Lett.\ {\bf 98}, 046402 (2007).}

\bibitem{VanderbiltReview}
	N.\ Marzari, A.\ A.\ Mostofi, J.\ R.\ Yates, I.\ Souza, and D.\ Vanderbilt,
	Rev. Mod. Phys. {\bf{ 84}}, 1419  (2012).
	
\bibitem{JanJensEmilSebastianPeter}
J.\ C.\ Budich, J.\ Eisert, E.\ J.\ Bergholtz, S.\ Diehl, and P.\ Zoller,
Phys. Rev. B {\bf 90}, 115110 (2014).	


\bibitem{JanSebastianTopDens}
J.\ C.\ Budich, S.\ Diehl, arXiv:1501.04135 (2015).


\bibitem{ReadGreen}
N.\ Read, D.\ Green,
Phys. Rev. B {\bf 61}, 10267 (2000). 



\bibitem{QAH}
\bibinfo{author}{{F.~D.~M. Haldane}},
  \bibinfo{journal}{Phys. Rev. Lett.} \textbf{\bibinfo{volume}{61}},
  \bibinfo{pages}{2015} (\bibinfo{year}{1988}).


\bibitem{DubailReadCriticalChern}
J.\ Dubail, N.\ Read, arXiv:1307.7726 (2013).

\end{thebibliography}

\end{document}